\documentclass[twocolumn]{aastex62}
\usepackage[utf8]{inputenc}
\usepackage{graphicx,natbib,color,amsmath,subfigure,comment}
\usepackage{hyperref}
\hypersetup{colorlinks=true,linkcolor=blue,urlcolor=blue,citecolor=blue,anchorcolor=blue}

\newcommand{\csf}{cold solar flare}
\newcommand{\Csf}{Cold solar flare}
\def\Mw{{Microwave}}
\def\mw{{microwave}}
\newcommand{\goes}{\textit{GOES}}
\newcommand{\kw}{Konus-\textit{Wind}}

\def\gyr{{gyrosynchrotron}}

\begin{document}

\title{Cold Solar Flares I. Microwave Domain}

\author[0000-0002-3942-8341]{Alexandra L. Lysenko}
\affiliation{Ioffe Institute, Polytekhnicheskaya, 26, St. Petersburg, 194021 -  Russian Federation}
\email{alexandra.lysenko@mail.ioffe.ru}

\author[0000-0002-8574-8629]{Stephen M. White}
\affiliation{Air Force Research Laboratory, Space Vehicles Directorate, Kirtland AFB, NM 87123, USA}

\author{Dmitry A. Zhdanov}
\affiliation{Institute of Solar-Terrestrial Physics  (ISZF), Lermontov st., 126a, Irkutsk, 664033 -  Russian Federation}

\author[0000-0002-6873-6394]{Nataliia S. Meshalkina}
\affiliation{Institute of Solar-Terrestrial Physics  (ISZF), Lermontov st., 126a, Irkutsk, 664033 -  Russian Federation}

\author[0000-0002-1589-556X]{Aleksander T. Altyntsev}
\affiliation{Institute of Solar-Terrestrial Physics  (ISZF), Lermontov st., 126a, Irkutsk, 664033 -  Russian Federation}

\author[0000-0001-7856-084X]{Galina G. Motorina}
\affiliation{Astronomical Institute of the Czech Academy of Sciences, 251 65 Ond\v{r}ejov, Czech Republic}
\affiliation{Central Astronomical Observatory at Pulkovo of Russian Academy of Sciences, St. Petersburg, 196140, Russia}

\author[0000-0001-5557-2100]{Gregory D. Fleishman}
\affiliation{New Jersey Institute of Technology, University Heights, Newark, NJ 07102-1982 -  USA}
\affiliation{Leibniz-Institut für Sonnenphysik (KIS), Freiburg, 79104, Germany}

\begin{abstract}
We identify a set of $\sim$100 ``cold'' solar flares and perform a statistical analysis of them in the microwave range. 
Cold flares are characterized by a weak thermal response relative to nonthermal emission. 
This work is a follow up of a previous statistical study of cold flares, which focused on hard X-ray emission to quantify the flare nonthermal component.  Here we focus on the microwave emission. 
The thermal response  is represented by the soft X-ray emission measured by the \textit{GOES} X-ray sensors. 
We obtain spectral parameters of the flare gyrosynchrotron emission and investigate patterns of the temporal evolution. 
The main results of the previous statistical study are confirmed: as compared to a ``mean'' flare, the cold flares have shorter durations,  higher spectral peak frequencies, and harder spectral indices above the spectral peak. 
Nonetheless, there are some cold flares with moderate and low peak frequencies. 
In a majority of cold flares, we find evidence suggesting the presence of the Razin effect in the microwave spectra, indicative of rather dense flaring loops. 
We discuss the results in the context of electron acceleration efficiency.
\end{abstract}

\section{\label{sec_intro}Introduction}
The magnetic energy that powers a solar flare can be distributed in a number of ways: heating of the surrounding plasma, accelerating charged particles (both electrons and ions), the kinetic energy of a coronal mass ejection, and radiation. 
The energy allocation between these different  components varies dramatically from flare to flare \citep{Emslie2012}, and it is yet unclear what causes the energy partitioning in each case. 

In purely ``thermal'' flares, detectable particle acceleration does not occur and almost all released magnetic energy is spent on direct plasma heating \citep[e. g.,][]{Gary1989, Fleishman2015}.
For the majority of flares, particle acceleration coexists with direct plasma heating \citep{Veronig2002}. 
For other flares, particle acceleration strongly dominates over direct heating, and almost all of the plasma thermal response is caused by the energy loss of the accelerated electrons. 
These flares are called ``cold'' solar flares and are characterized by a weak thermal response  relative to the nonthermal emission. 
For \csf s no significant thermal emission is observed prior to the impulsive phase where nonthermal particles dominate, thus they represent a subclass of ``early impulsive flares'' \citep{Sui2006}. 


\Csf s are of particular interest. 
First, they allow experimental exploration of the causes of energy partitioning in a solar flare. 
Second, in such flares emission from nonthermal electrons can be examined down to low ($\sim$10\,keV) energies without contamination from stronger thermal emission. 
Third, \csf s allow us to study the thermal response of the plasma to accelerated particles without the admixture of direct heating.

Several \csf s were reported in previous case studies \citep{White1992, Bastian2007, Fleishman2011, Masuda2013, Fleishman2016, Motorina2020}. 
The energy balance between thermal and nonthermal components calculated for cold flares in \cite{Bastian2007}, \cite{Fleishman2016} and \cite{Motorina2020} confirmed that the energy conveyed by accelerated electrons was sufficient to explain the observed plasma heating. 
The weak thermal response was attributed to (a) low temperatures due to high plasma density, in the flares reported by \cite{Bastian2007} and \cite{Masuda2013}; or (b) low emission measure due to low plasma density \citep{Fleishman2011} or the small volume of the main flaring loop \citep{Fleishman2016}.

In \cite{Lysenko2018} (hereafter L18) a statistical study of \csf s in the X-ray and microwave ranges was performed for the first time. 
Cold flares were selected using hard X-ray (HXR) data from the \kw\ experiment \citep{Aptekar1995}, complemented by the \mw\ data from several available instruments. The reference flare set in the HXR range has been selected from solar flares registered by \kw\ and not classified as cold flares. The reference flare set in the \mw\ range was taken from the statistical study by \cite{Nita2004} based on observations at the Owens Valley Solar Array (OVSA). 

L18 found that, by comparison with the reference groups, cold flares tend to be characterized by harder spectral indices of the accelerated electron energy spectra. 
In the microwave range the overall peak frequency distribution of \gyr\ spectra for cold flares was significantly shifted towards higher frequencies as compared to the reference flare set. 
At the same time, there are cold flares with very low peak frequencies. 
Compared to the reference groups of flares, cold flares are characterized by shorter duration in both the \mw\ and HXR ranges. 
Thus, L18 concluded that a group of cold flares characterized by high peak frequencies is associated with compact loops with strong magnetic fields, while the low-frequency group is produced by larger loops with weak field. 
It remains a question whether the harder spectra of accelerated electrons are related to the acceleration mechanism involved in cold flares, or whether chromospheric evaporation is reduced due to the penetration of electrons with harder spectra into the deeper layers of the Sun's atmosphere \citep{Fisher1985, Reep2015}.

In the present statistical study we use \mw\ emission for the selection of \csf s instead of HXR emission in order to provide a different perspective on this phenomenon. 
The aims of the research are to cross-check the results with L18; to explore if the selection criteria for \csf s are sensitive to the choice of the nonthermal emission regime, HXR or \mw ; which results are resistant to the flare selection criteria and which are not. 
The present study explores the thermal response of the ambient plasma to accelerated electrons, flare evolution in \mw\ and X-ray ranges, and draws conclusions about flare morphology and flare properties related to the acceleration mechanism. 
A practical goal of the research is to extend our list of well-observed cold flares for future case studies. 

This paper is the first part of the research and is focused on event selection and the analysis of cold flare properties in the \mw\ domain. 
In the forthcoming second part we will study X-ray emission of \csf s and the relationships between properties observed in the \mw\ and X-ray domains.

\section{\label{sec_instr}Instrumentation}

\subsection{\label{ssec_instr_mw}Total Power Radio Instruments}

Key radio instruments used in this work are the Nobeyama Radiopolarimeters \citep[NoRP,][]{Torii_etal_1979} located in Japan.
NoRP measures intensity and circular polarization at six frequencies (1, 2, 3.75, 9.4, 17, and 35\,GHz) with a time resolution of 1\,s  along with intensity measurements only at 80\,GHz.
In addition to NoRP, we use observations from several other radio instruments: the US Air Force Radio Solar Telescope Network  \citep[RSTN,][]{Guidice1981}, the Solar Radio Spectropolarimeters \citep[SRS,][]{Muratov2011}, and the Badary Broadband Microwave Spectropolarimeters \citep[BBMS,][]{Zhdanov2015}. 
RSTN consists of four stations at Learmonth (Australia), San Vito (Italy), Sagamore Hill (USA) and Palehua (USA), which measure intensity at eight frequencies (245, 410, 610, 1415, 2695, 4995, 8800, and 15400\,MHz) with a 1\,s time resolution. 
In this work we use observations by Learmonth and Palehua stations overlapping in time with NoRP.
SRS and BBMS spectropolarimeters are located near Irkutsk, Russia and provide integrated flux over the whole solar disk in two circular polarizations. 
SRS covers the 2–24 GHz frequency range at 16 frequencies with a temporal resolution of 1.6\,s.
BBMS performs measurements in the range 4--8\,GHz at 26 frequencies with resolution of 10\,ms.

\subsection{\label{ssec_instr_sxr} \textit{GOES} data in Soft X-ray Range}

In the soft X-ray (SXR) range we use the data from the \textit{GOES} X-ray sensors (XRS) in two broad bands, 1--8\,\AA\ and 0.5--4\,\AA\ \citep{White2005}.
Spacecraft of NOAA's \textit{GOES} series have been performing SXR observations since 1974 with temporal resolution varying from 3\,s to 1\,s. It should be noted that the XRS flux scale was changed with the transition to GOES-16 in 2020\footnote{see \url{http://www.ngdc.noaa.gov/stp/satellite/goes/doc/GOES_XRS_readme.pdf}}, and NOAA are in the process of converting older data to the modern scaling, but most existing flare catalogs use the legacy scaling and that is the scale used for XRS data here.


\subsection{\label{ssec_instr_image}Imaging instruments}

When available, the locations of \csf s on the solar disk were determined by \mw\ images from Nobeyama Radioheliograph \citep[NoRH,][]{Nakajima1994}.
In other cases we searched for HXR images from the Ramaty High Energy Solar Spectroscopic Imager \citep[\textit{RHESSI},][]{Lin2002}. 
For the cases where neither NoRH nor \textit{RHESSI} data were available,  we used differential images obtained by Atmospheric Imaging Assembly onboard Solar Dynamics Observatory  \citep[\textit{SDO}/AIA,][]{Lemen2012}.\footnote{The  \textit{SDO}/AIA images are available on the website \url{https://www.lmsal.com/solarsoft/latest_events_archive/events_summary/} or \url{https://helioviewer.ias.u-psud.fr/}.}

\section{\label{sec_cold_selection}Selection of Cold Solar Flares}

\begin{figure}
    \centering
    \includegraphics[width=0.47\textwidth]{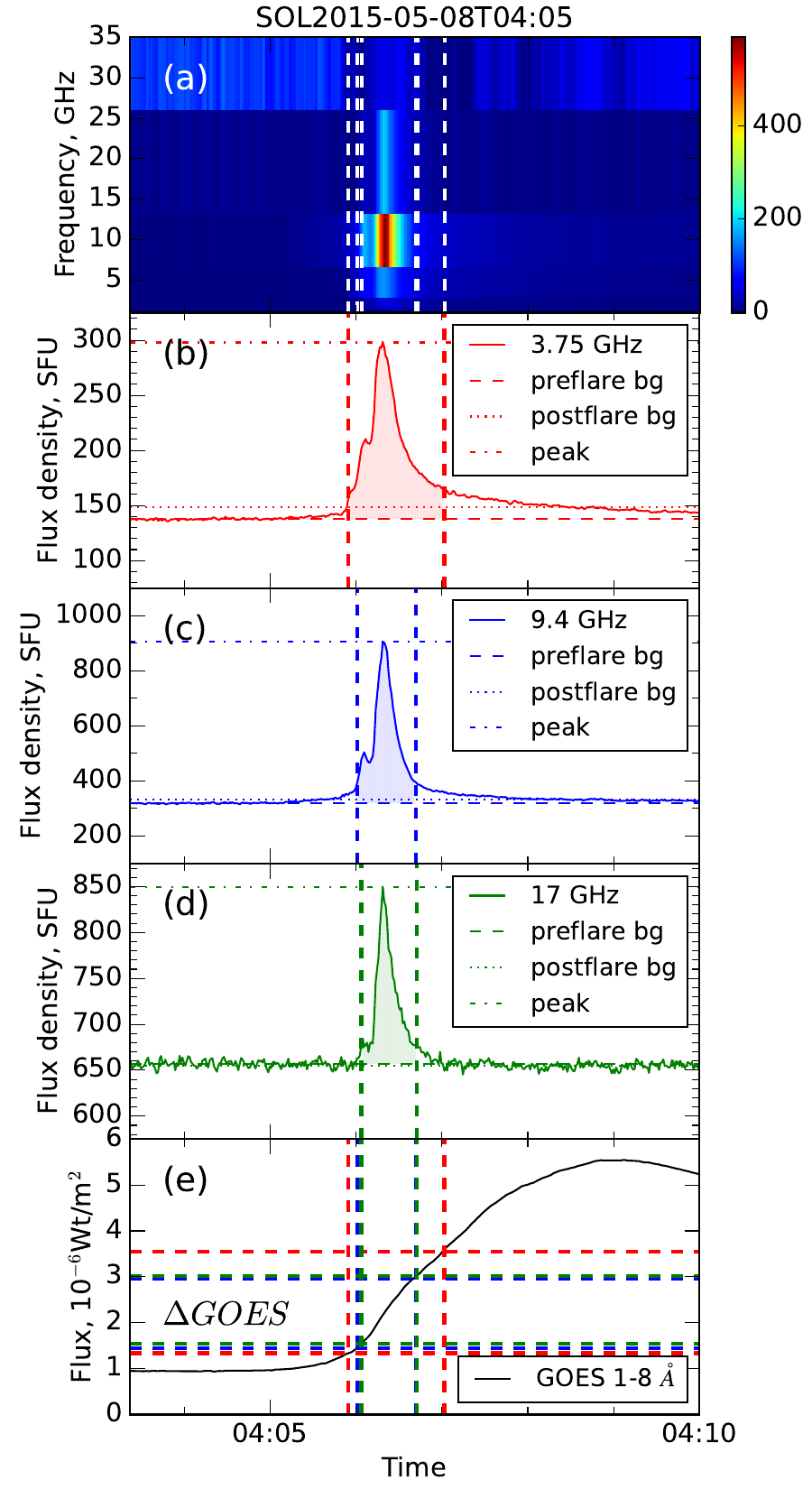}
    \caption{Association between thermal and nonthermal flare components. 
    (a) \Mw\ dynamic spectrum of a flare on 2015 May 8; the colorbar represents the flux density in SFU; time profiles at (b) 3.75\,GHz, (c) 9.4\,GHz, and (d) 17\,GHz; horizontal dashed lines correspond to the preflare background level, dotted lines indicate the postflare background level, dash-dotted lines mark flare peaks at each frequency, semitransparent fill corresponds to time-integrated flux density at each frequency; (e) flux in the \goes\ 1--8\,\AA\ channel, horizontal dashed lines mark the flux increment $\Delta GOES$ during the impulsive phase at 3.75\,GHz (red), 9.4\,GHz (blue), 17\,GHz (green);  vertical dashed lines indicate the beginning and the end of the flare impulsive phase at 3.75\,GHz (red), 9.4\,GHz (blue), 17\,GHz (green).}
    \label{fig_norp_goes_ex}
\end{figure}

\begin{figure*}
    \centering
    \includegraphics[width=0.9\textwidth]{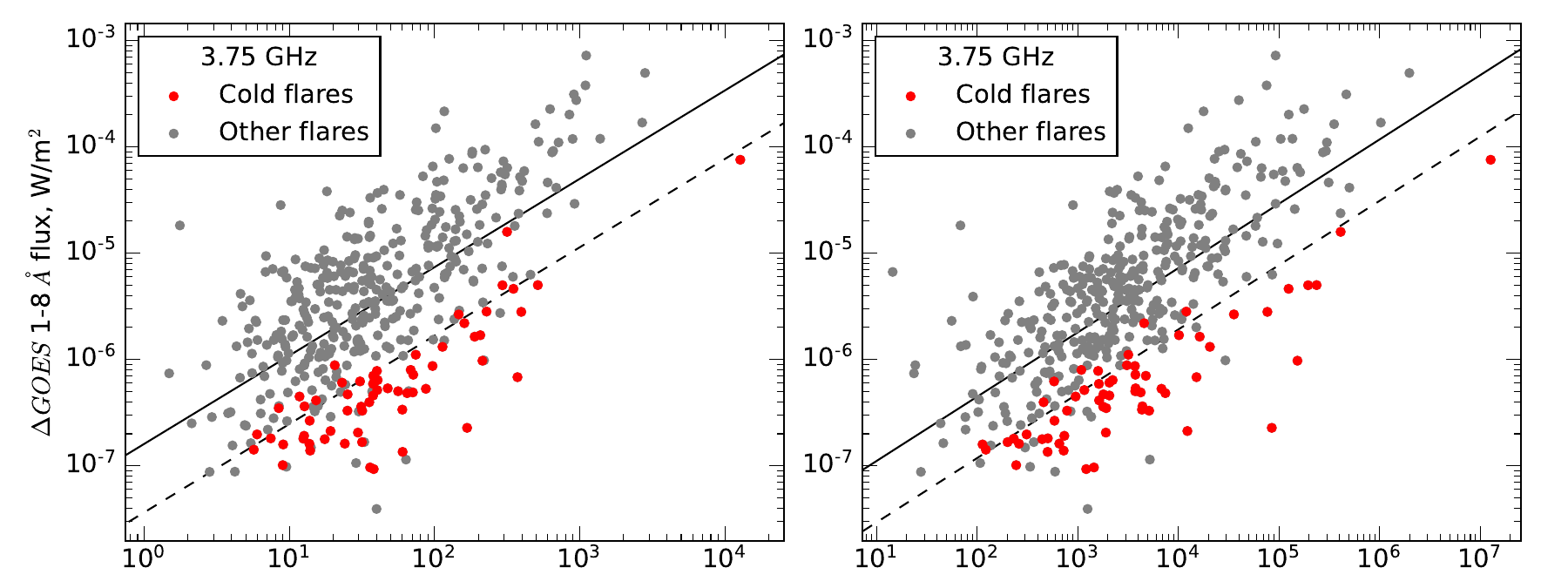} \\
    \includegraphics[width=0.9\textwidth]{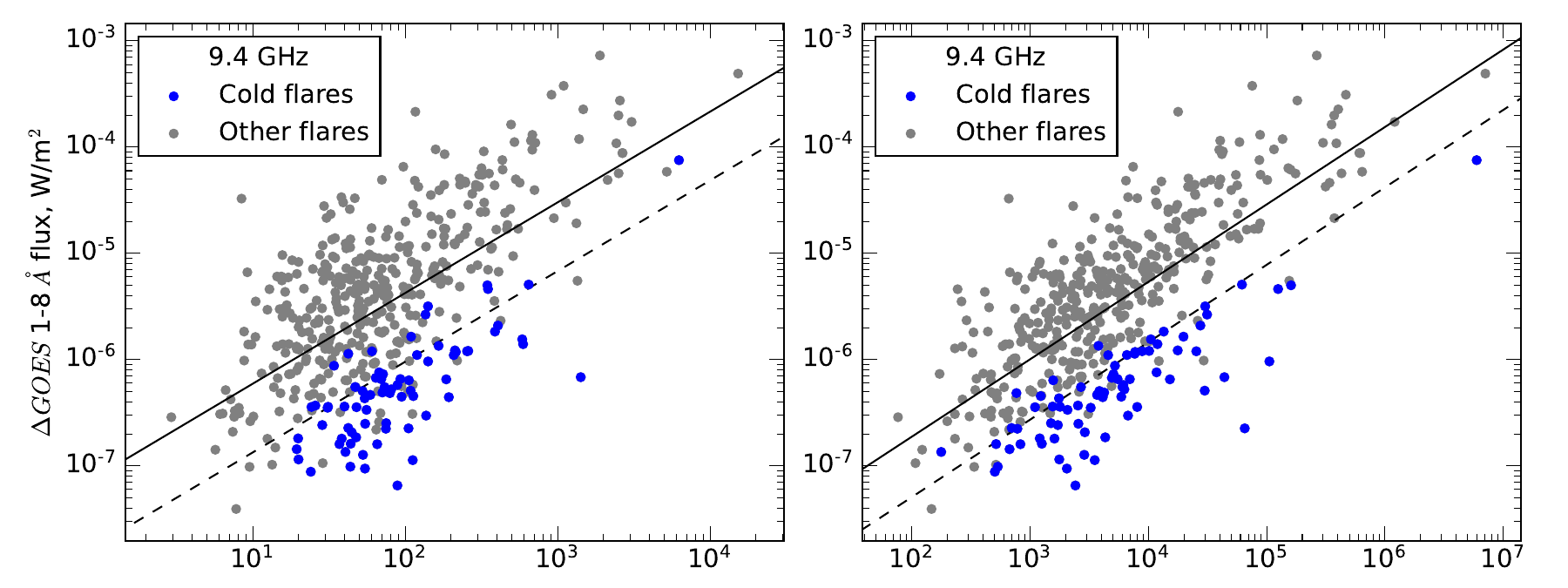} \\
    \includegraphics[width=0.9\textwidth]{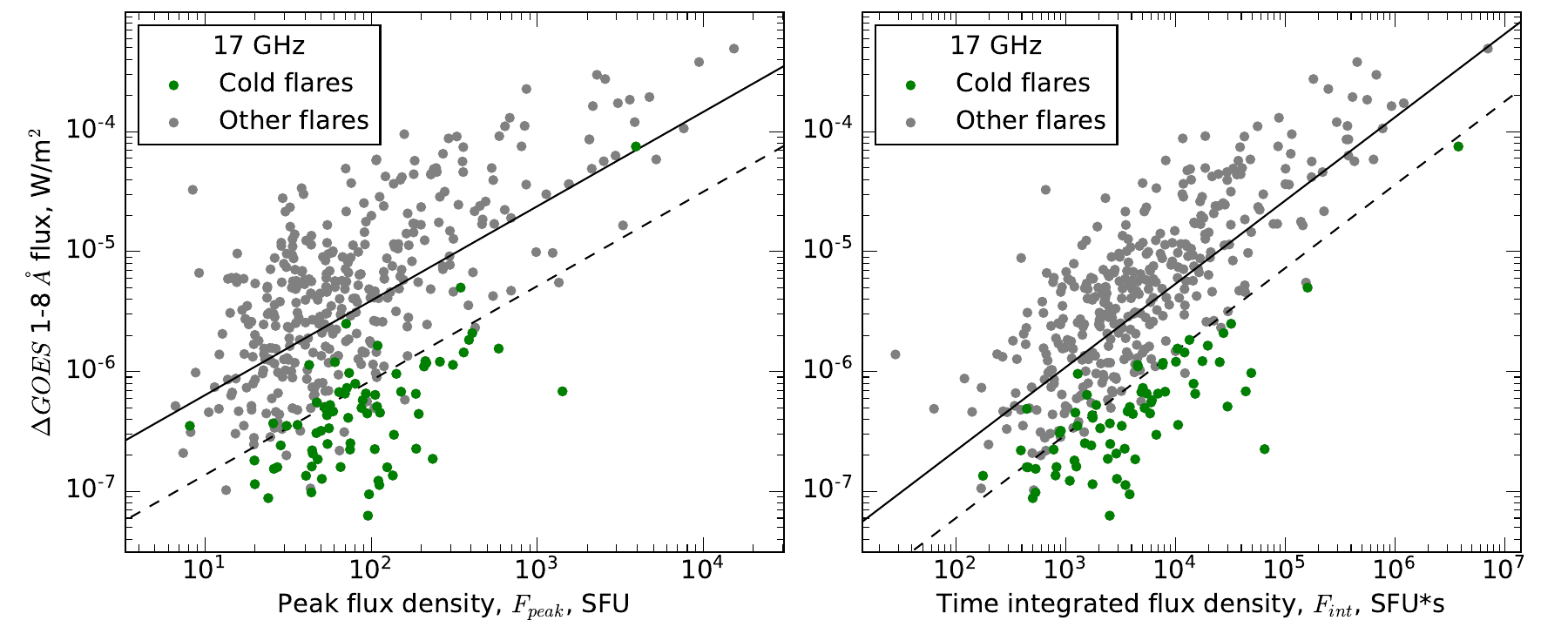} 
    \caption{Relationship between peak flux density (left) and time integrated flux density (right) observed by NoRP and the flux increment in the \goes\ 1--8\,\AA\ channel during the flare impulsive phase at 3.75\,GHz (top), 9.4\,GHz (middle) and 17\,GHz (bottom). 
    Solid black lines are regression lines, dashed black lines represent the bounds which distinguish the cold flares. 
    Cold flares are represented as red (3.75\,GHz), blue (9.4\,GHz) and green (17\,GHz) filled circles, other flares are marked as grey circles.
    }
    \label{fig_norp_vs_goes}
\end{figure*}

We searched for \csf s in the events seen by NoRP from 2010 to 2017 and identified matching observations by \textit{SDO}/AIA  and \textit{RHESSI} for future case studies of the selected flares. 
For this search we used a new catalog of NoRP events developed for other purposes (a study of the incidence of coherent emission; S. M. White et al., in preparation). 
This catalog is somewhat larger than the observatory catalog, which stopped updating in 2015 due to resource issues: the new catalog has 633 events in the period 2010-2015 relevant for this study, compared to 195 in the on-line catalog. 


For  comparison between thermal and nonthermal components we used an approach similar to that adopted in L18. 
The nonthermal component was quantified as the background-subtracted peak flux density or time-integrated flux density of \gyr\ emission observed in the \mw\ range during the impulsive phase. 
The thermal response to this nonthermal energy input was estimated as the maximum flux increase in the \goes\ 1--8\,\AA\ channel during the flare impulsive phase, $\Delta GOES$ (Figure~\ref{fig_norp_goes_ex}). 
If no response was observed, or a high and decreasing background in the \goes\ 1--8\,\AA\ channel was present, $\Delta GOES$ was estimated as the flux error at the beginning of the impulsive phase in the 1--8\,\AA\ channel, i.e., 15\,\% of the flux \citep{Garcia1994}. 
In the cases where thermal response was very low, i. e. less than the error of the maximum flux, we also estimated upper limits for $\Delta GOES$ as 15\,\% of the maximum flux.

L18 identified two main groups in the radio properties of \csf s: high-frequency flares with a maximum in the radio spectrum at frequencies $>$10\,GHz, and low-frequency flares with spectral maximum at a few GHz. 
Along with these main groups, there are cold flares with spectral peaks at moderate frequencies (between $\sim$3 and $\sim$10\,GHz). 
To account for this spectral diversity we performed a search on three NoRP frequencies: 3.75\,GHz, 9.4\,GHz and 17\,GHz. 
We did not use 1\,GHz and 2\,GHz for the search because of the frequent occurrence of bright coherent emission there.
At higher frequencies we considered narrow-band events with a short time scale to be caused by coherent emission and excluded them.

A bias could well result from the use of multiple frequencies to pick out cold flares without accounting for the location of the radio spectral peak. 
This is because the \mw\ flux depends on a high power of the magnetic field strength as well as the nonthermal energy of the radio-emitting electrons, and thus the pure radio flux is, likely, not the best measure of the nonthermal energy: the higher the spectral peak, likely the higher the magnetic field strength and thus the more the radio flux might over-represent the nonthermal energy. 
On the other hand, regions of strong magnetic field tend to be more compact than regions with weaker field. 
The radio flux is proportional to the source area when optically thick; thus, the radio flux increase due to magnetic field increase may be partly compensated by a corresponding decrease of the source area. 
We therefore check \textit{a posteriori} if our selection introduces any bias towards events with higher \mw\ spectral peak frequency.


There is an intrinsic uncertainty in determining the end of the impulsive phase in the \mw\ range because of thermal radio emission at later stages. 
Often, the flare-associated thermal radio emission decays very slowly and lasts for minutes or tens of minutes after the \gyr\ emission is over, similar to the duration of SXR emission. 
To identify and exclude this component we segmented the \mw\ time profiles into ``Bayesian blocks'', which is a standard technique for astronomical light-curve analysis \citep{Scargle2013}.
A Bayesian block is a time interval assumed to have ``constant'' flux density at the selected significance level, with superposed variations that can be regarded as random fluctuations.
Here we took the significance level to be equal to 4$\sigma$ where $\sigma$ was determined for each flare on each of the three frequencies as a standard deviation of the flux density during a 30\,s interval with stable emission. 
We determined preflare and postflare background levels as the first Bayesian block with duration more than 30\,s before and after the flare peak respectively.\footnote{We used our own implementation for Bayesian block segmentation. The source code is available via \url{https://github.com/dsvinkin/b_blocks}} 
The beginning and the end of the impulsive phase were determined as the times when the flux is at a level 10\,\% of the peak value above preflare and postflare background levels, respectively (Figure~\ref{fig_norp_goes_ex}). 
Time profiles at frequencies with failures, significantly varying background, and significant contribution from coherent emission were excluded.

Finally we derived power-law relations between the peak flux density and $\Delta GOES$ and time-integrated flux density and $\Delta GOES$ for more than 500 flares. 
These data are plotted in Figure~\ref{fig_norp_vs_goes}. 
Solid lines represent linear regression lines (on a logarithmic scale) for all flares between the peak flux density (Figure~\ref{fig_norp_vs_goes}, left) and the time-integrated flux density (Figure~\ref{fig_norp_vs_goes}, right) at  3.75\,GHz (top), 9.4\,GHz (middle) and 17\,GHz (bottom) and $\Delta GOES$. 
Regression lines (solid black) were calculated using the python \textsf{scipy.stats.linregress} procedure. 
Dashed lines separate outliers with low thermal response at each of the three frequencies.
These outliers were selected as in L18: we built the distribution of distances between each point and the regression line for the specified frequency, with negative distances for the points below and positive distances for the points above the regression line. 
The points which fall in the 16\,\% percentile are identified as weak thermal response outliers.
In this manner we selected 130 flares which demonstrated weak thermal response relative to the nonthermal peak or time-integrated emission on at least one of the three frequencies. 

From this candidate list we excluded flares which are not early impulsive flares, i.e. flares with preheating.
We used the formal criterion proposed in \cite{Sui2006}: the absence of a flux increase in the \goes\ XRS sensors earlier than 30\,s before the flare impulsive phase. 
Here we met with a difficulty -- almost half of the cold flare candidates occurred during unstable  background in the SXR range, either during the decay of a more powerful flare, or after a series of small brightenings, perhaps, preflares. 
Such preflares are rather common phenomena \citep[see, e.g.][]{Farnik1996, Farnik1998, Battaglia2009, Wang2017}, and for the majority of flares the preflare sources don't coincide spatially with the main flare phase \citep{Farnik1998} and refer to a distinct energy release such as reconnection of a small loop with a larger one \citep{Farnik1996, Wang2017}. 
Without spatial information it is difficult to draw a solid conclusion as to whether activity observed before the flare refers to the same reconnection act as the impulsive phase and, thus, may represent a preheating, or not. 
Thus, we used the following formal criteria.
If the flux in the \goes\ sensors increases monotonically before the impulsive phase, we considered it to be preheating and excluded such flares. 
If before the impulsive phase the SXR flux increased and than decreased, we kept such flares as \csf\ candidates. 
These will be further discussed in Section~\ref{sec_disc}.

These criteria identified 109 early impulsive solar flares without preheating and with weak thermal response. 
Later we excluded two flares after cross-calibration of NoRP data with BBMS and SRS, which revealed that the flux density in the \mw\ range for these flares showed significant discrepancies between observatories. 
Localization of these 107 flares on the solar disk revealed two behind-the-limb flares which were also excluded since the thermal emission could, in fact, be strong but occulted by the solar limb. 

The final list contains 105 \csf s. 
Figures and tables with the relationship between nonthermal and thermal flare components for each cold flare are reported at doi:\href{https://doi.org/10.5281/zenodo.7775771}{10.5281/zenodo.7775771}. 
Cold flares are plotted in Figure~\ref{fig_norp_vs_goes} with red, blue and green circles referring to frequencies 3.75\,GHz, 9.4\,GHz and 17\,GHz, respectively, while the remaining flares are shown as gray circles. 
As we used weak thermal response relative to either peak or time-integrated nonthermal emission, there are cold flares above the dashed lines on all relationships in Figure~\ref{fig_norp_vs_goes}.

The cold flare list presented here contains 15 solar flares registered by \kw\ in the triggered mode. 
The \csf\ list from L18 contains six cold flares which coincide in time with NoRP observations and occur in 2010--2017. 
Thus, the cold flare selection in L18 was more restrictive and there are only four flares in common between these two lists. 

\section{\label{sec_analysis}Data Analysis}

\subsection{\label{ssec_analysis_loc}Distribution of \csf s on the disk}

\begin{figure}
    \centering
    \includegraphics[width=0.47\textwidth]{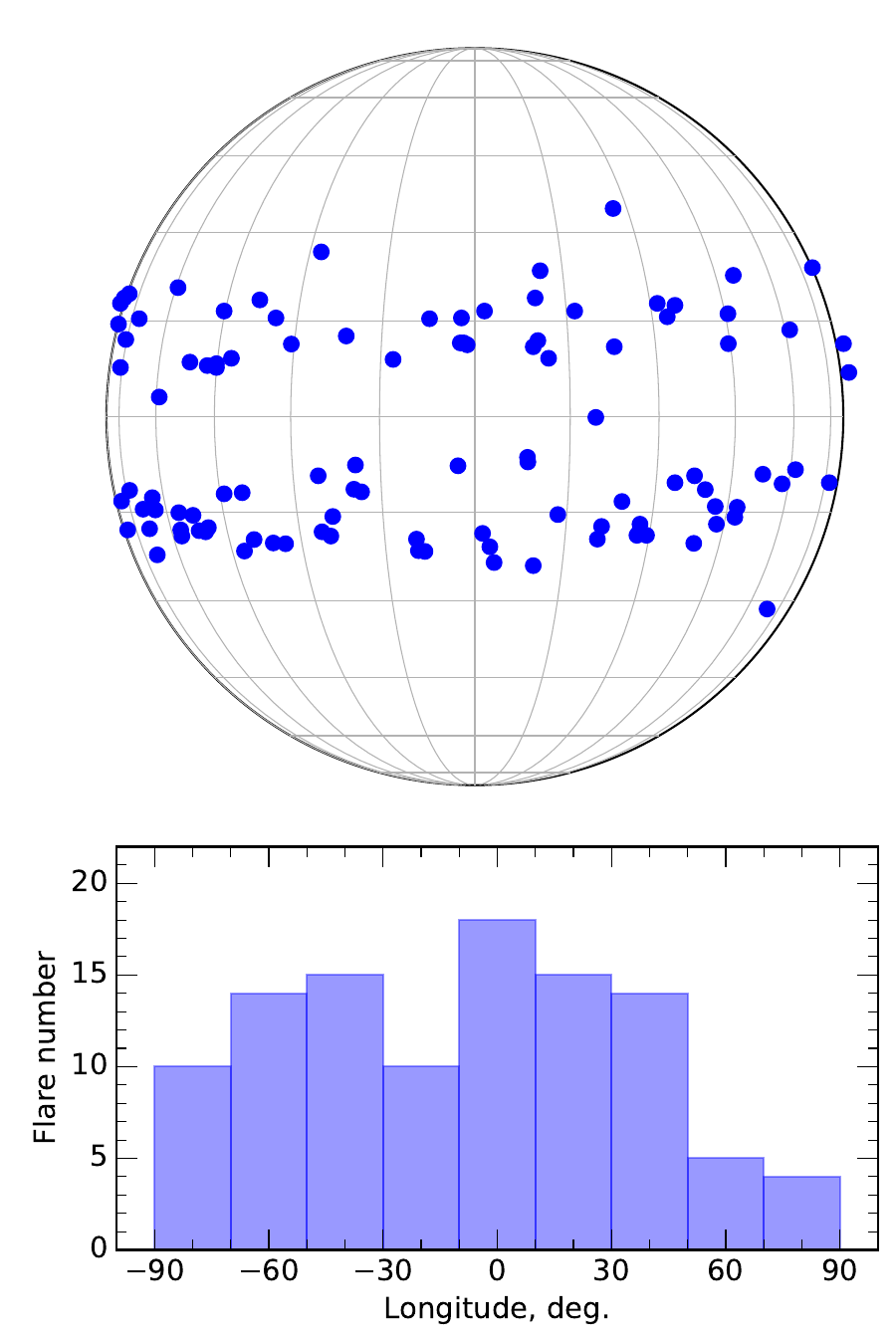}
    \caption{\Csf\ locations on the solar disk (top) and distribution by longitude (bottom). }
    \label{fig_loc}
\end{figure}

\Csf\ locations on the solar disk are plotted in Figure~\ref{fig_loc}, along with their distribution with longitude. 
Helioprojective coordinates for each cold flare are reported at doi:\href{https://doi.org/10.5281/zenodo.7775771}{10.5281/zenodo.7775771}.
The flare distribution  over longitude is fairly uniform taking into account the uncertainties.
\cite{Kosugi1985} showed that solar \gyr\ emission at 17\,GHz is apparently isotropic, and the slight bias away from the limb obtained by him is due to the difficulty of near-limb flare identification in $H_{\alpha}$ images.  
Thus, we conclude that the \csf\ longitudial distribution is similar to that of other flares, and the selection of \csf s is not affected by observational selection related to emission directivity. 

\subsection{\label{ssec_analysis_mw_intro}Spectral fitting of \gyr\ \mw\ spectra}

\begin{figure}
    \centering
    \includegraphics[width=0.48\textwidth]{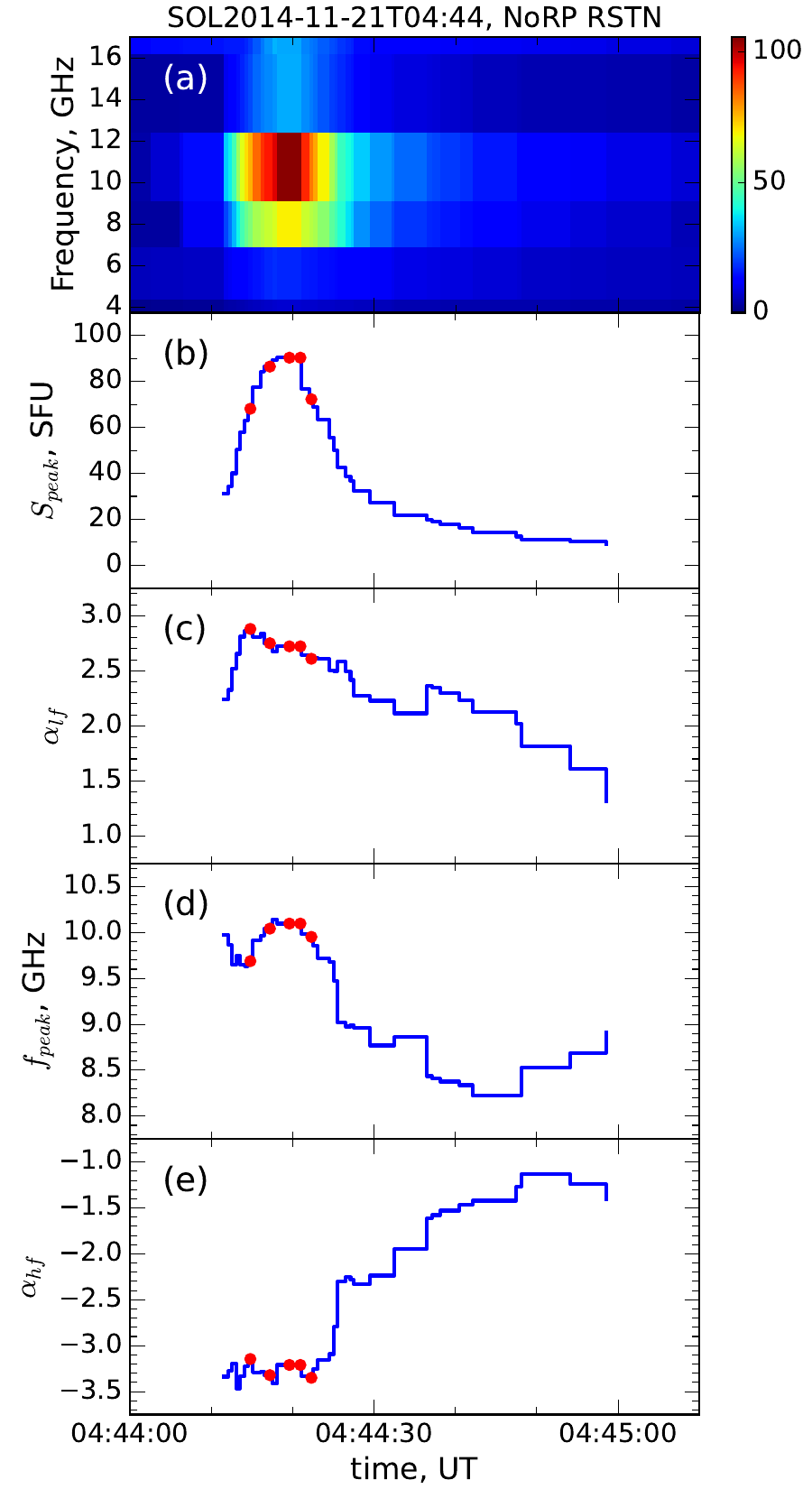}
    \caption{An example of fitting a \csf\ with the GYR model (equation 1; cf. L18). 
    (a) \Mw\ dynamic spectrum; the color bar shows the flux density scale in SFU; time evolution of (b) the peak flux density; (c) the low frequency spectral index; (d) the peak frequency; and (e) the high frequency spectral index. 
    Red points mark time intervals included in histograms with fitting parameters. }
    \label{fig_mw_fit}
\end{figure}

The \gyr\ spectrum of flare-accelerated electrons depends on the magnetic field strength and direction, the spectrum of nonthermal electrons, and the density and the temperature of the thermal plasma, see online video in \citet{Fleishman2022}. It has a bell-shape that can  be characterized by four free parameters: the peak frequency $f_{\rm peak}$, the peak flux density $S_{\rm peak}$, the low-frequency spectral index $\alpha_{\rm lf}$, and the high-frequency spectral index $\alpha_{\rm hf}$. 
This spectral shape can be described by a phenomenological model \citep{Stahli1989}:

\begin{equation}
\label{eq_mw_fit}
S(f)=A f^{\alpha}\left[1-e^{-{B} f^{-\beta}}\right],
\end{equation}
Here $\alpha_{\rm lf}$=$\alpha$, $\alpha_{\rm hf}$=$\alpha$-$\beta$ and $f_{\rm peak}$ and $S_{\rm peak}$ can be calculated via parameters of the function $S(f)$.

To determine parameter values at a given time frame we use least-square  spectral fitting. 
Prior to the spectral fitting, we prepared dynamic spectra combining all available \mw\ data  obtained with various instruments. 
We fixed clock errors by selecting NoRP timing as the reference. 
Time intervals for the background estimation were defined individually for each instrument, and background was approximated by a constant or a polynomial.
The standard deviation at each frequency was estimated as the flux density deviation from the background level during a background time interval.
To reduce random fluctuations of the fitting parameters we used Bayesian block segmentation (see Section~\ref{sec_cold_selection}). 
The time intervals used for the spectral fitting were combined from the boundaries of Bayesian blocks at each frequency, thus, fitting was performed on the data with variable time resolution. 
We ignored frequencies with faults, amplitude calibration errors and frequencies where coherent plasma emission could be present. 
If there were multiple spectral peaks we analyzed the most intense one.

The least-squares technique was applied in python using the \textsf{scipy.optimize.minimize} function. 
Wherever possible we fitted the spectra by the phenomenological \gyr\ model (GYR, Equation~\ref{eq_mw_fit}), but for a number of flares the spectral peak was outside the observed frequency range. 
In these cases we used a power-law model (PL) and determined only lower limits for $S_{\rm peak}$, lower or upper limits for $f_{\rm peak}$ and $\alpha_{\rm lf}$ or $\alpha_{\rm hf}$ depending on whether $f_{\rm peak}$ is above or below the observational range, respectively. 
For two flares we had to freeze $\alpha_{\rm lf}$ at fixed values in order to achieve a fit. 
For some flares $\alpha_{\rm lf}$ or $\alpha_{\rm hf}$ was poorly defined when using the GYR model due to the small number of observed frequencies below or above the spectral peak. In these cases the appropriate spectral index was defined using the PL model. 
For four flares both $\alpha_{\rm lf}$ and $\alpha_{\rm hf}$ were rather large such that the GYR model couldn't determine them correctly, and the PL model was used for estimation of both spectral indices. 

We obtained successful fits for 86 flares with the GYR model and for 12 flares with the PL model. 
For the remaining seven flares spectral fitting failed due to insufficient number of observed frequencies or large amplitude-calibration discrepancies between different instruments. 
An example of the time evolution of the fitted parameters is plotted in Figure~\ref{fig_mw_fit}.
Fit results for individual \csf s are plotted at doi:\href{https://doi.org/10.5281/zenodo.7775771}{10.5281/zenodo.7775771}

\subsection{\label{ssec_analysis_mw_dist}Parameter distributions of \gyr\ spectra }
\begin{figure*}
    \centering
    \includegraphics[width=0.9\textwidth]{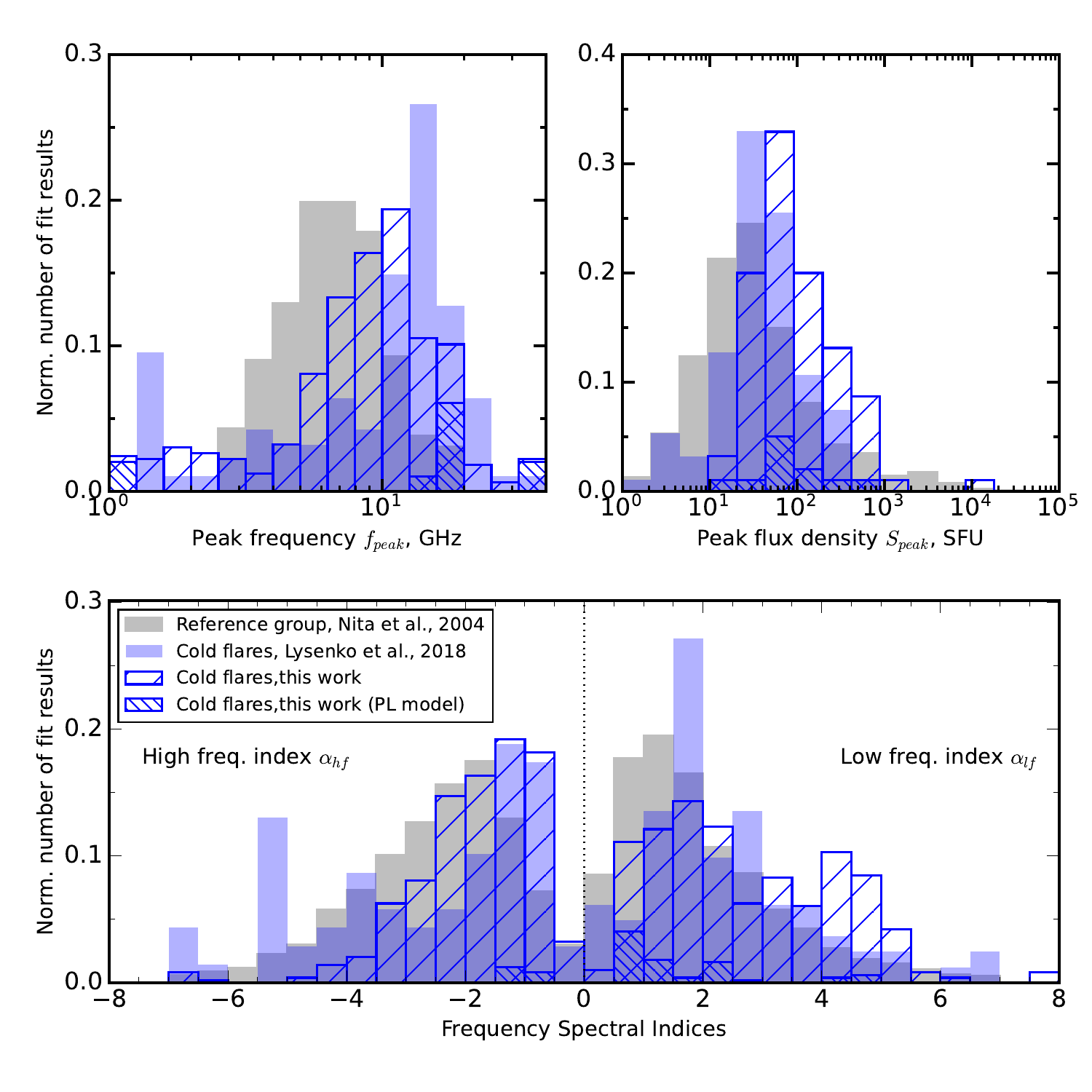}
    \caption{Distributions of the \mw\ spectral parameters for \csf s in the present study (blue hatching), \csf s from L18 (light blue), reference flare set (gray). 
    Inverse hatching denotes cold flares from this work fitted with the PL model. 
    Top left: peak frequency distribution; top right: peak flux density distribution; bottom: distribution of spectral indices in the low frequency range (right) and in the high frequency range (left).}
    \label{fig_mw_specpars}
\end{figure*}

Following L18, we compare distributions of the fitted parameters $S_{\rm peak}$, $f_{\rm peak}$, $\alpha_{\rm lf}$ and $\alpha_{\rm hf}$ obtained for \csf s with those from a group of reference flares obtained by \cite{Nita2004} based on OVSA data \citep{Gary1994}. 
Unlike L18 and the present study, the reference group contains events with both \gyr\ and coherent emission at the lower frequencies. 
To account for this slight discrepancy we included in the reference distributions only flares with peak frequencies above 2.6\,GHz; this frequency was found to be a rough demarcation point between decimetric, often coherent, and centimetric, mainly \gyr\ bursts.

To account for significant time evolution of spectral parameters which can occur during the \mw\ bursts \citep[see, e.g.,][]{Melnikov2008, Fleishman2016} we used the same approach as in L18: histograms with parameter distributions include values obtained for five time frames during the main temporal peak. 
These five time frames are selected at the beginning and the end of the burst, at the peak maximum, and in the middle of the rising and declining phases. 
The beginning and the end of the burst were defined in the same way as in \cite{Nita2004}: as the beginning and the end of the time interval during which the flux density at the peak frequency of the event is above 80\,\% of the corresponding peak flux. 
The parameter values included in histograms are marked in Figure~\ref{fig_mw_fit} with red points.

Distributions of the peak flux density $S_{\rm peak}$, peak frequency $f_{\rm peak}$, low frequency spectral index $\alpha_{\rm lf}$ and high frequency spectral index $\alpha_{\rm hf}$ are presented in Figure~\ref{fig_mw_specpars}, along with results obtained for \csf s in L18 and results for the reference group of solar flares taken from \cite{Nita2004}. 

The peak frequency distribution for \csf s obtained in this work is shifted to higher frequencies as compared to the distribution of the reference flare set, but this shift is not as large as the one obtained in L18. 
This could be evidence that our multi-frequency \csf\ selection process does not introduce additional bias towards events with high spectral peak frequency compared with the L18 selection based on the HXR data (see Section~\ref{sec_cold_selection}).
As in L18, there is a group of \csf s with low peak frequencies ($f_{\rm peak}<$3\,GHz), i. e. 14 flares out of 98 flares with successful fits; 37 flares are characterized by moderate values of $f_{\rm peak}$ (3\,GHz$<f_{\rm peak}<$10\,GHz); while the most numerous group among \csf s, 47 out of 98, are high-frequency flares with $f_{\rm peak}>$10\,GHz. 
Higher values of peak frequency indicate stronger magnetic field in the \csf s compared to flares from the reference group \citep{Fleishman2020}.

The distribution of the peak flux density shows that the \csf s studied in this work are more intense than those studied in L18 and those from the reference group. 
This can be attributed to harder electron energy spectra and, as a consequence, higher effective energies of the radiating nonthermal electrons for \csf s as well as to higher magnetic field.

The low-frequency spectral index could be determined for 96 \csf s out of 98 flares with successful fits. 
\Csf s studied in this work are characterized by higher values of $\alpha_{\rm lf}$ than the \csf s from L18 and flares from the reference set. 
The magnitude of $\alpha_{\rm lf}$ is related to the source morphology \citep{Fleishman2018}: 
lower values of $\alpha_{\rm lf}$ can be explained by source inhomogeneity, while larger values of $\alpha_{\rm lf}$ can be related to, e.g., Razin supression \citep{Razin1960a}, free-free absorption, or \gyr\ self-absorption along the line of sight \citep{Bastian1998}, for further details see Section~\ref{ssec_analysis_mw_ev}. 

The high-frequency spectral indices $\alpha_{\rm hf}$ for \csf s both from this work and L18 are smaller in absolute value than those for the  reference flare set obtained by \cite{Nita2004}.
There could be several reasons for this; e.g., harder energy spectra of accelerated electrons \citep{Fleishman2020} or the presence of the Razin effect which could flatten $\alpha_{\rm hf}$ \citep{Razin1960b, Melnikov2008}.
It should be noted that the percentage of flares with very flat spectra ($\alpha_{\rm hf}>$-0.5) presumably related to free-free emission of thermal electrons \citep{White1992} is approximately the same for \csf s from this work, from L18 and for the reference flare set.

\subsection{\label{ssec_analysis_mw_dur}Timescales in the \mw\ range}

\begin{figure}
    \centering
    \includegraphics[width=0.47\textwidth]{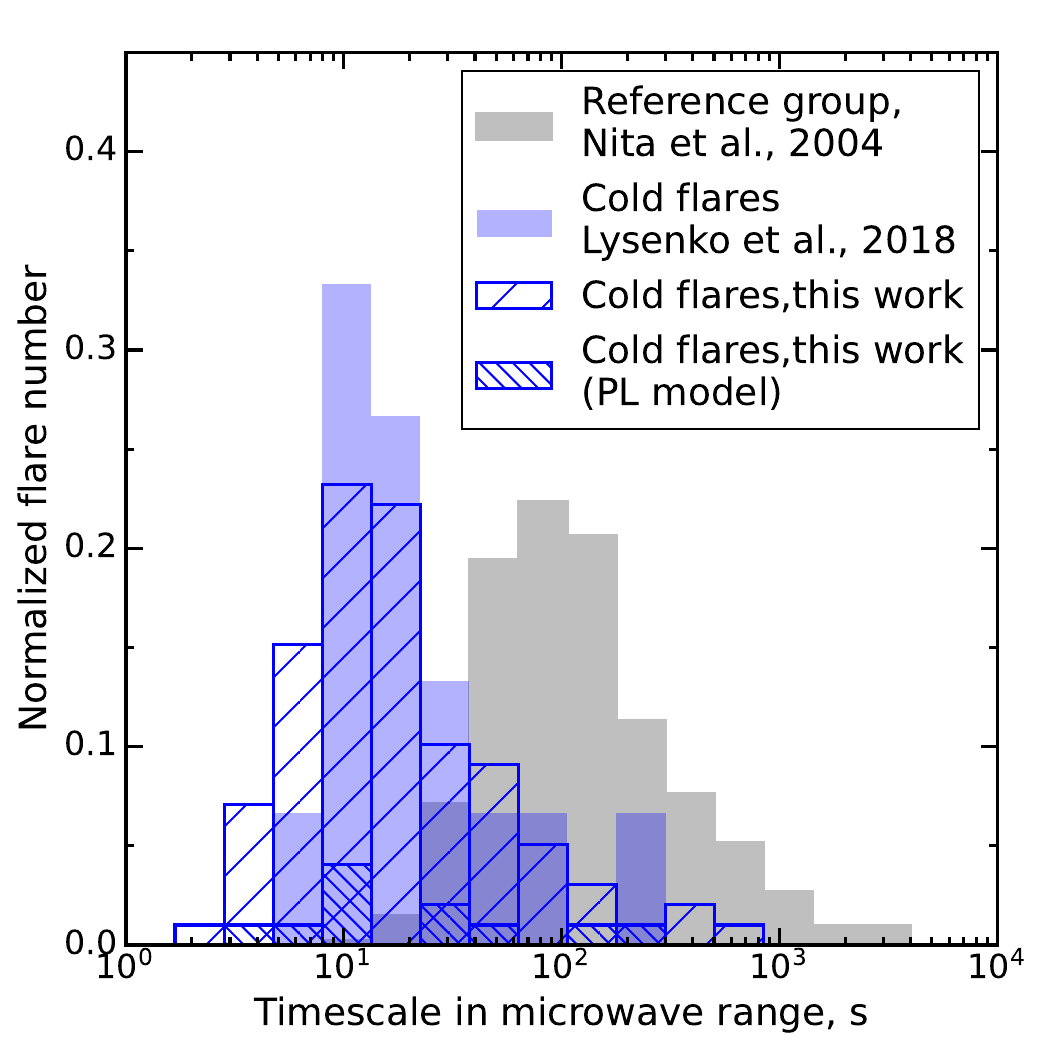}
    \caption{Distribution of the timescales in the \mw\ range for \csf s from the presented study (blue hatching), from L18 (light blue) and for the reference flare group (gray). 
    Inverse hatch denotes cold flares from this work fitted by the PL model. }
    \label{fig_mw_dur}
\end{figure}

\Csf\ timescales in the \mw\ range were estimated as the time interval during which the flux density at the peak frequency exceeds 80\,\% of the peak flux (see Section~\ref{ssec_analysis_mw_dist}). 
Such an approach allows direct comparison of the present results with the results obtained in L18 and in \cite{Nita2004}, but it can only be applied to the flares with successful spectral fits. For the flares fitted by the PL model a lower limit of the timescale is defined. 
Note that the timescale defined in this way is significantly shorter than the impulsive phase duration defined from the light curves (Section~\ref{sec_cold_selection}).
The peak timescales for individual flares are reported at doi:\href{https://doi.org/10.5281/zenodo.7775771}{10.5281/zenodo.7775771}, and their distribution along with distributions for the \csf s from L18 and for the reference flare set from \cite{Nita2004} are presented in Figure~\ref{fig_mw_dur}. 
\Csf s both from this work and L18 are generally significantly shorter than flares from the reference group. 
This could imply shorter flare loops for the \csf s compared to the flares from the reference group.

\subsection{\label{ssec_analysis_mw_ev}Spectral evolution in the \mw\ range}

\begin{figure*}
    \centering
    \includegraphics[width=0.46\textwidth]{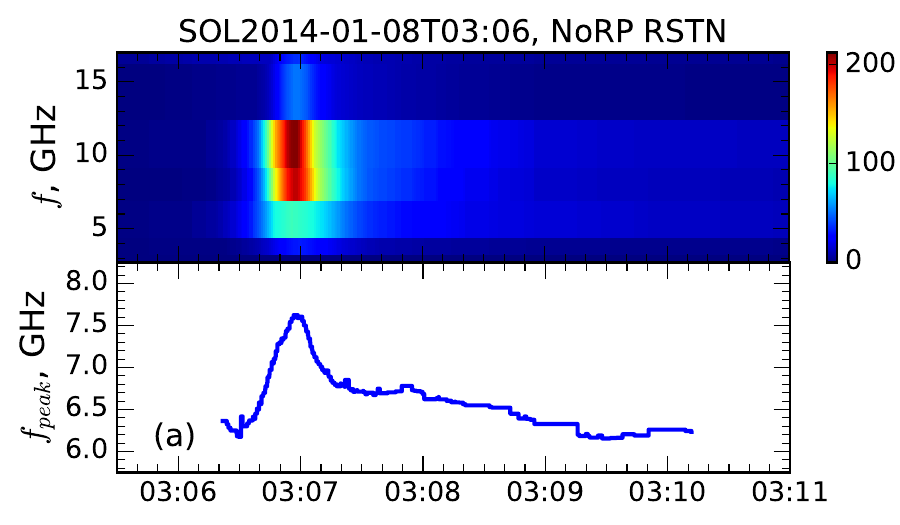}
    \includegraphics[width=0.46\textwidth]{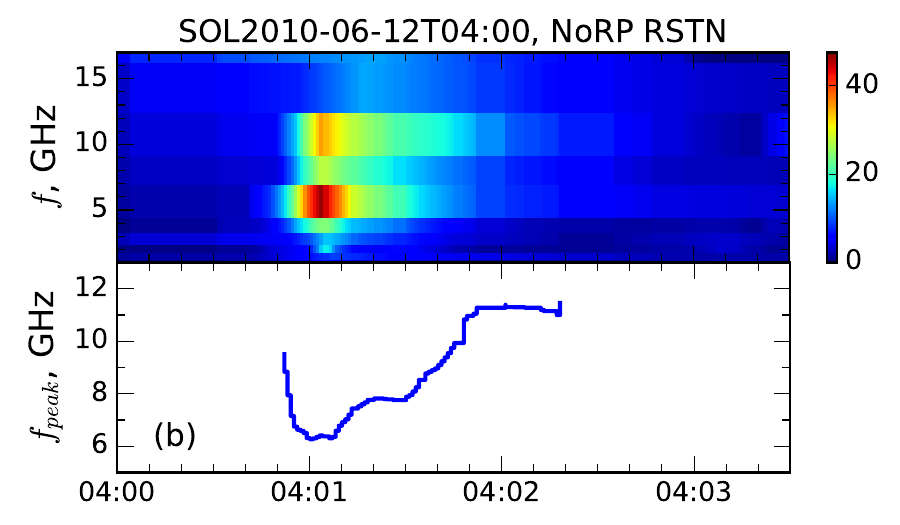} \\
    \includegraphics[width=0.46\textwidth]{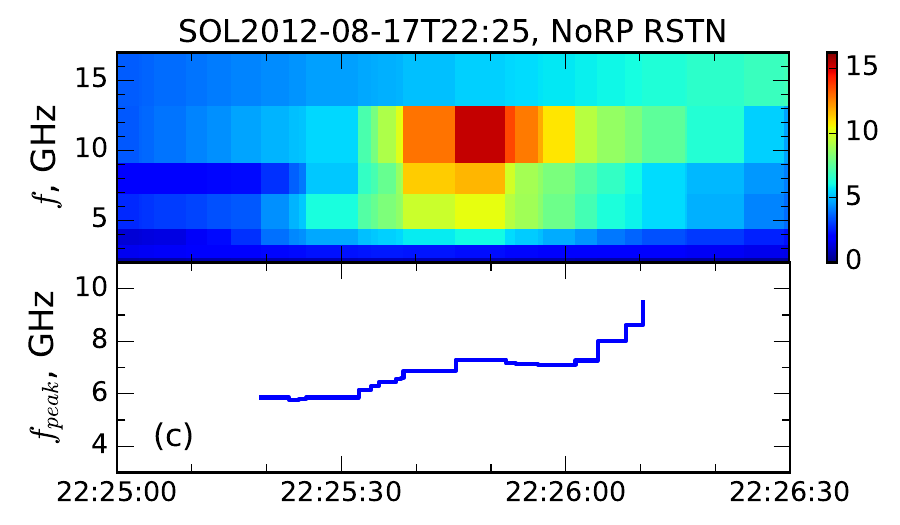}
    \includegraphics[width=0.46\textwidth]{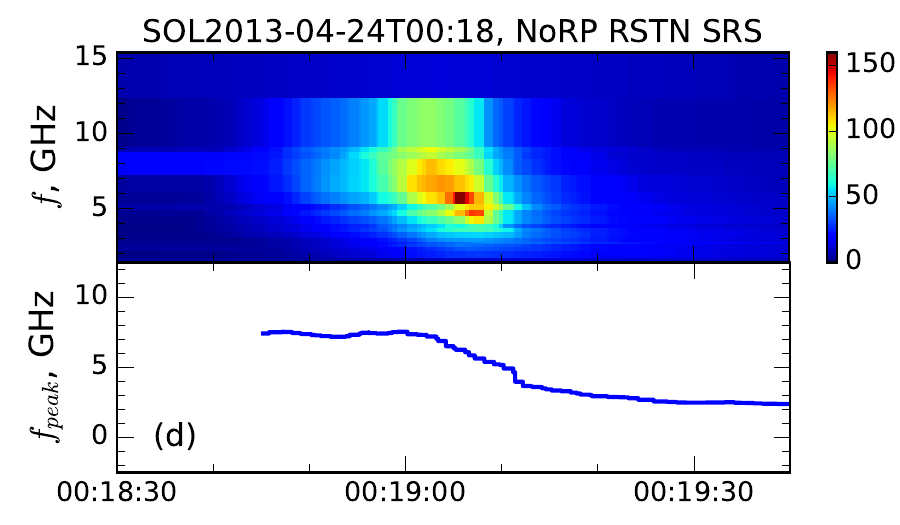} \\
    \includegraphics[width=0.46\textwidth]{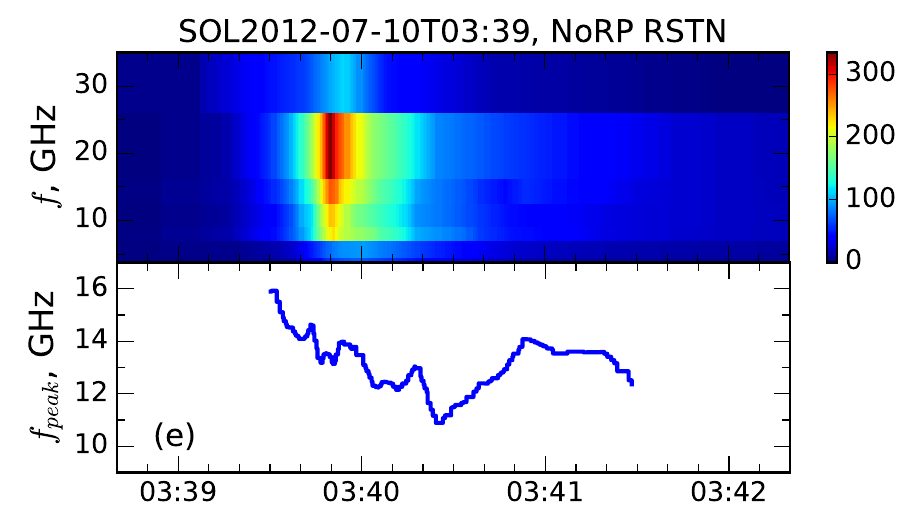}
    \includegraphics[width=0.46\textwidth]{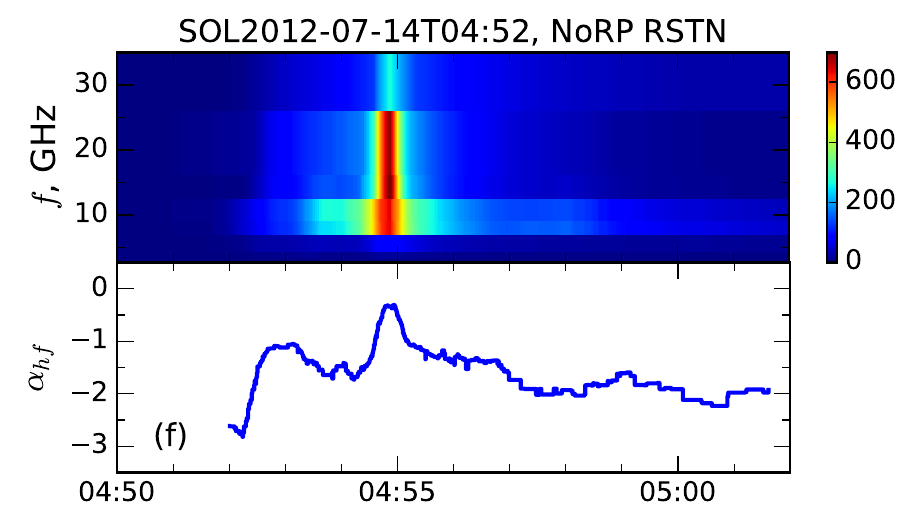} \\
    \includegraphics[width=0.46\textwidth]{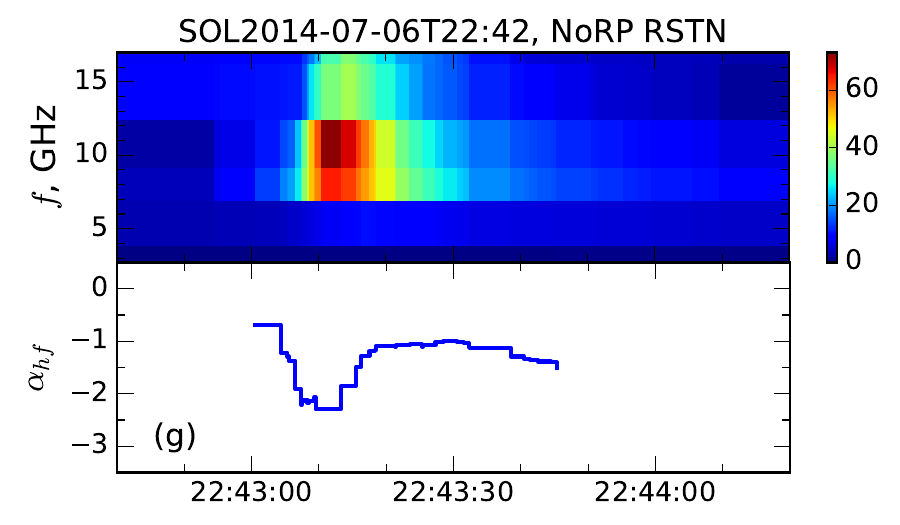} 
    \includegraphics[width=0.46\textwidth]{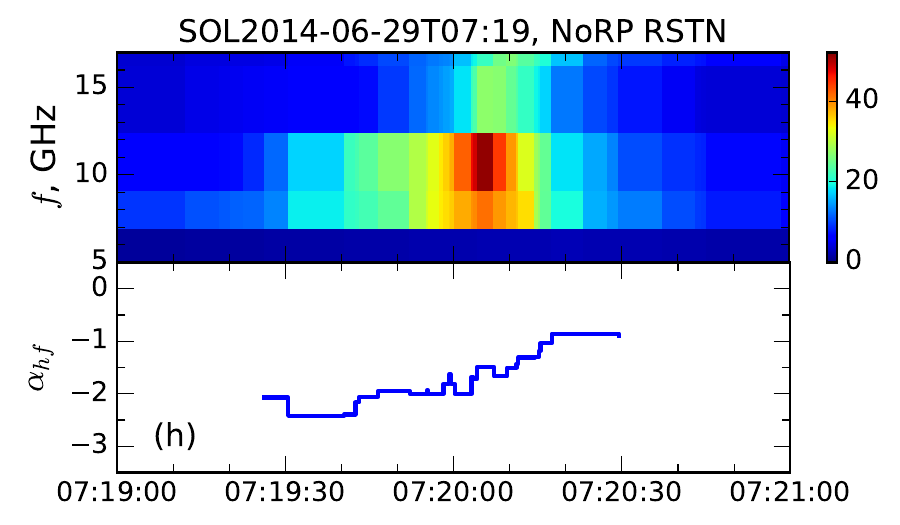} \\
    \caption{Examples of \mw\ spectral parameter evolution. 
    (a) Correlation between peak frequency $f_{\rm peak}$ and peak flux density $S_{\rm peak}$, ``C+''; (b) anticorrelation between $f_{\rm peak}$ and $S_{\rm peak}$, ``C--''; (c) increase of $f_{\rm peak}$ during the flare, type ''F+''; (d) decrease of $f_{\rm peak}$ during the flare, ``F--''; (e) multiple increases/decreases of $f_{\rm peak}$ during the flare main peak, ``F?''; (f) soft-hard-soft evolution of high frequency index $\alpha_{\rm hf}$, SHS; (g) hard-soft-hard evolution of $\alpha_{\rm hf}$, HSH; (h) soft-hard-harder evolution of $\alpha_{\rm hf}$, SHH. 
    }
    \label{fig_mw_ev}
\end{figure*}

Evolution of the \mw\ spectral parameters is linked with the evolution of flare morphology and the nonthermal electron distribution, and, thus,  can help better understand dynamic processes in solar flares  \citep[e.g.][]{Bastian2007, Melnikov2008, Fleishman2016}. 
The peak frequency evolution is available for 86 out of 98 flares via successful fits with the GYR model (see Section~\ref{ssec_analysis_mw_intro}). 
We consider two aspects of $f_{\rm peak}$ variations: correlation with the peak flux density, and the overall change in peak frequency over the course of the flare. 

Based on the correlations between $f_{\rm peak}$ and $S_{\rm peak}$, the flares were divided into three groups. 
The first group, ``C+'', demonstrates high correlation (correlation coefficient $C_{\rm fS}>$0.5) and contains 34 flares; an example of such a flare is presented in Figure~\ref{fig_mw_ev}(a). 
In the second group, ``C--'', high anticorrelation between $f_{\rm peak}$ and $S_{\rm peak}$ is observed ($C_{\rm fS}<$-0.5); this group contains 16 flares (Figure~\ref{fig_mw_ev}(b)). 
Flares of the third group, ``C0'', do not show a significant relationship between $f_{\rm peak}$ and $S_{\rm peak}$ (-0.5$<C_{\rm fS}<$0.5, 36 flares). 
High correlation between $f_{\rm peak}$ and $S_{\rm peak}$ implies that the spectral peak is formed by \gyr\ self-absorption (GS absorption), while deviations from this relation most likely indicate the presence of the Razin effect \citep{Melnikov2008}. 
As the Razin cutoff frequency is proportional to $f_{\rm R}\sim n_e/B$, where $n_e$ is the number density of electrons in the ambient plasma and B is the magnetic field, high anticorrelation between $f_{\rm peak}$ and $S_{\rm peak}$ could be explained as the increase of the magnetic field strength simultaneously with the flux density. 
An alternative explanation could be significant radio source expansion while maintaining the same number of emitting particles \citep{Lee1994, Fleishman2016}.

We identified four patterns of overall peak frequency change during the flare: overall increase of $f_{\rm peak}$ (``F+'', 27 flares (an example is presented in Figure~\ref{fig_mw_ev}(c)); overall decrease of $f_{\rm peak}$ (``F--'', 33 flares, Figure~\ref{fig_mw_ev}(d)); the recovery of $f_{\rm peak}$ after the main peak to the level observed before the peak (``F0'', 13 flares); and more complicated evolution with multiple increases and decreases of $f_{\rm peak}$ during the main flare peak (``F?'', 13 flares, Figure~\ref{fig_mw_ev}(e)). 
The increase of the peak frequency could be related to the increase of plasma density during the flare in the presence of the Razin effect \citep{Melnikov2008}.
The peak frequency decrease during the flare was observed for the \csf\ described by \cite{Bastian2007} and was explained as  flare-loop heating by accelerated electrons, making the loop more transparent to the free-free emission. 
Another explanation of the $f_{\rm peak}$ decrease could be the movement of the \mw\ source to the loop top where the magnetic field is weaker \citep{Lee1994}.
Multiple peak frequency variations observed for the flares of type ``F?'' could be caused by the variations of the source size, viewing angle and/or number density of nonthermal electrons during the main flare peak \citep{Fleishman2020}. 

A homogeneous source with a spectral peak formed by \gyr\ absorption should have a low-frequency index between 2 and 3, while the Razin effect for a homogeneous source produces spectra with steeper low-frequency  spectral indices \citep{Razin1960b}. 
Free-free absorption by a thermal plasma between the source and the observer can result in even steeper low-frequency spectral slopes \citep{Bastian1998}. 
An accurate analysis in \cite{Fleishman2018} based on multifrequency imaging showed that the influence of the source inhomogeneity can reduce low-frequency indices in the presence of the Razin effect down to values of $\sim$1. 

In the present study we could not perform such a thorough analysis, thus, for the rough flare classification we employed formal criteria based on $\alpha_{\rm lf}$ values and the relationship between $f_{peak}$ and $S_{peak}$.
Events with $\alpha_{\rm lf}<$2 and high correlation between $f_{\rm peak}$ and $S_{\rm peak}$ were attributed to inhomogeneous sources (I -- 21 flares); the combination of $\alpha_{\rm lf}<$4 and low correlation between $f_{\rm peak}$ and $S_{\rm peak}$ was interpreted as inhomogeneous sources with the Razin effect (IR -- 28 flares); flare spectra which showed correlation between $f_{\rm peak}$ and $S_{\rm peak}$ were assumed to be formed by GS absorption if 2$<\alpha_{\rm lf}<$3 (S -- 6 flares), or GS absorption plus free-free absorption along the line of sight if 3$<\alpha_{\rm lf}<$4 (SA -- 7 flares); cases with 4$<\alpha_{\rm lf}<$6 were interpreted as the Razin effect in homogeneous sources (R -- 31 flares); high values of the low-frequency index, $\alpha_{\rm lf}>$6, were explained by a combination of Razin effect and free-free absorption along the line-of sight (RA -- 3 flares).
Thus the Razin effect likely played a role on the low-frequency side for the majority of flares, 62 out of 96 flares with estimations of $\alpha_{\rm lf}$.

High-frequency indices are primarily determined by the energy spectra of accelerated electrons in flare loops, but may be affected by other factors as well. 
These factors include pitch-angle anisotropy of the nonthermal electrons, the strength of the magnetic field, the Razin effect, or contribution from the thermal free-free component. 
Accordingly, the evolutionary patterns of the high-frequency spectral index may show more diversity than do the HXR indices \citep[this is confirmed by the catalog of events employed by][who, however, did not report detailed statistical properties of this spectral evolution]{Nita2004}. 

We found three main classes for the evolution of $\alpha_{\rm hf}$.
The first type, soft-hard-soft (SHS) evolution, is typical for hard X-ray spectra and is observed for 25 \csf s; an example is shown in Figure~\ref{fig_mw_ev}(f). 
This evolution most likely mirrors the properties of the electron acceleration mechanism \citep{Grigis2004}. 
However, the most numerous group consists of 40 flares with the opposite, hard-soft-hard behavior of $\alpha_{\rm hf}$ (HSH, see Figure~\ref{fig_mw_ev}(g)). 
Such behavior could be attributed to the contribution of free-free emission with a flatter spectrum over the course of the flare. 
Another explanation is the change of the electron pitch-angle distribution from isotropic to beam-like while observing from a transverse direction during the flare peak \citep{Fleishman2003}.
Another flare group demonstrates soft-hard-harder (SHH) spectral evolution, and contains 10 flares (Figure ~\ref{fig_mw_ev}(h)). 
This evolution type is rather common for both HXR and microwave emission \citep{Asai2013} and could be explained by the capture of accelerated electrons in magnetic traps and preferential scattering of low energy electrons into the loss cone due to Coulomb collisions leading to their precipitation into the chromosphere. This results in a steady hardening of the energy spectrum of the electrons remaining in the trap \citep{Cliver1986}. 
For the remaining flares we could not draw solid conclusions about the evolution of $\alpha_{hf}$ due to the insufficient number of frequencies above the spectral peak, or chaotic changes in $\alpha_{hf}$ making it difficult to determine a clear evolution pattern. 

The patterns of parameter temporal evolution along with morphology types defined by $\alpha_{\rm lf}$ for individual cold flare are reported at doi:\href{https://doi.org/10.5281/zenodo.7775771}{10.5281/zenodo.7775771}.

\section{\label{sec_disc} Summary and Discussion}
In the present study we have analyzed \mw\ properties of solar flares with abnormally weak thermal response relative to the nonthermal emission---\csf s.
The selection criteria for \csf s were similar to those used in 
L18: (i) a low flux increase in the $GOES$ 1--8\,\AA\ channel during the flare impulsive phase relative to the peak flux or time integrated flux of the nonthermal emission; and (ii) the absence of  preheating before the flare impulsive phase. 
In this study we used \mw\ emission at 3.75\,GHz, 9.4\,GHz and 17\,GHz for the estimation of the nonthermal flare component, rather than HXR emission as used in L18. 
Our search revealed 105 \csf s out of a subset of $>$500 solar radio bursts recorded by NoRP in the period 2010--2017, hence  $\sim$20\,\% of flares were qualified as ``cold.'' 
In L18, only 27 \csf s were found from 1994 to 2017 among $\sim$1000 solar flares registered by \kw\ in the triggered mode, thus the criteria used in the current work are less restrictive. 
This is partly because \kw\ detects only rather hard and spiky flares in the triggered mode, while NoRP data are less exposed to these selection effects. 
Thus, as compared to L18, the present flare list improves the statistical sample for the study. 

Despite the differences in the choice of the nonthermal diagnostic, the main results of L18 for the parameter distribution of the \gyr\ spectra were confirmed here. 
As compared to the reference flare set taken from \cite{Nita2004}, the \csf s have the following features:

\begin{enumerate}
\item Cold flares both from L18 and the present work are characterized by higher values of the \gyr\ peak frequencies $f_{\rm peak}$: in the present study $\sim$50\,\% of cold flares have $f_{\rm peak}>$10\,GHz, while statistical studies of \mw\ bursts showed that for the majority of bursts $f_{\rm peak}$ lies in the range 4--10\,GHz \citep{Guidice1975, Stahli1989, Nita2004}.

\item Both studies revealed a group of flares with low peak frequencies ($f_{\rm peak}<$3\,GHz); in the present study this group contains $\sim$14\,\% of cold flares. 

\item Distribution of the low-frequency indices $\alpha_{\rm lf}$ for \csf s is shifted towards higher values as compared to this distribution for the flares of reference group; although, this shift is much more vivid in the present study than in L18.

\item \Csf s from present work and L18 are characterized by harder high-frequency spectral indices $\alpha_{\rm hf}$ compared with the reference group.

\item \Csf s in present study demonstrate higher \mw\ intensities compared to cold flares from L18 and the reference set.

\item Cold flares from both this work and L18 have significantly shorter timescales in the \mw\ range, than flares from the reference set.

\end{enumerate}

In the present study, in addition to the distributions of \gyr\ spectral parameters, we examined their temporal evolution and found several evolutionary patterns allowing us to draw additional conclusions about flare morphology and flare scenarios.

If the spectral peak is formed by \gyr\ self-absorption, the peak frequency $f_{\rm peak}$ increases together with the peak flux density $S_{\rm peak}$ during the flare maximum \citep{Melnikov2008, Fleishman2020}, as the number of nonthermal electrons increases. 
However, for the majority of \csf s ($\sim$65\,\%), $f_{\rm peak}$ and $S_{\rm peak}$ do not correlate. 
This fact along with higher values of $\alpha_{\rm lf}$ speaks in favor of the Razin effect playing a role \citep{Melnikov2008}.
For $\sim$20\,\% of the \csf s the influence of the Razin effect is so strong that the $f_{\rm peak}$ and $S_{\rm peak}$ even anticorrelate: as the Razin cutoff frequency is inversely proportional to the magnetic field strength, such anticorrelation could be caused by an increase in field strength during the flare peak (for alternative explanations, see Section~\ref{ssec_analysis_mw_ev}). 

In many cases (one third of all \csf s) the peak frequency decreased during the course of the flare, which was also observed in the cold flare described by \cite{Bastian2007}. 
Presumably, in such flares accelerated electrons lose their energy in dense flaring loops before they reach the chromosphere. 
The heating of the loop by nonthermal electrons reduces the free-free opacity, which may reduce the peak frequency. 


For the group of flares with a spectral peak formed by \gyr\ self-absorption, high values of $f_{\rm peak}$ imply a strong magnetic field, and low values of $f_{\rm peak}$ are associated with a weak field. 
The presence of the Razin effect indicates high background density in the flaring loops; although the Razin cutoff frequency is inversely proportional to magnetic field, the field needs to be strong enough to produce intense \mw\ emission. 
The shorter durations of \csf s relative to the flares of the reference set could be explained by shorter flaring loops. 
Thus, many of the \csf s are inferred to be confined flares with high magnetic field strength and high density of the loop plasma. 
For such dense flares, chromospheric evaporation could be suppressed as most accelerated electrons lose their energy in the flaring loops and do not reach the chromosphere.
Such dense loops could be leftovers of some earlier activity (see Section~\ref{sec_cold_selection}), such as an earlier flare, which could have initiated significant chromospheric evaporation prior to the impulsive phase of the \csf\ \citep{Battaglia2009}. 
Between \csf s there is a group of flares with a spectral peak formed by \gyr\ self-absorption and low peak frequencies and, hence, weak field. 
Such flares could be similar to the tenuous cold flare described by \cite{Fleishman2011}, where the thermal response is reduced due to low emission measure. 

Recent simulations \citep{Arnold2021, Sioulas2022} show that the key parameter defining the efficiency of electron acceleration is the ratio between reconnecting and non-reconnecting magnetic field, or ``guide'' field. 
With the increase of the reconnecting field over the guide field, and, thus, the increase of the free magnetic energy, acceleration efficiency goes up.
Higher values of reconnecting magnetic field relative to the guide field give harder power-law indices for the electron distribution  \citep{Arnold2021}, which can be related to the harder high-frequency indices observed for \csf s compared to the flares of the reference group. 

\cite{Arnold2021} also show that system size has little influence on the nonthermal electron production, thus \csf s could be examples of high acceleration efficiency, similar to that
observed in the large X8.2 class solar flare of 10 September 2017 \citep{Fleishman2022}. 
In this flare almost all electrons in the cusp region were accelerated to form a power-law distribution, while the number density of the thermal electrons became undetectable.  

While the source of electron nonthermal energy gain in a flare is the magnetic field, the only driving force capable of accelerating charged particles is the electric field force. 
Data-constrained 3D modeling of a multi-loop C-class solar flare evolution (Fleishman et al. 2023; ApJ in press) revealed strong dependence of the acceleration efficiency on the Dreicer field value in the given flux tube. 
Specifically, the acceleration was highly efficient in a hot tenuous loop with a low value of the Dreicer field, while it was much weaker in a cooler and denser loop with a larger Dreicer field. 
This implies that the acceleration efficiency can be controlled by a balance between the effective electric field (which can be either turbulent, or regular, or a combination of both) responsible for plasma energization and the value of the Dreicer field that controls the fraction of the runaway particles vs the Maxwellian core particles.
In this context high acceleration efficiency in the \csf s with the apparent lack of the direct heating can indicate that the effective accelerating electric field is comparable to or larger than the Dreicer field. 
The Dreicer field depends on the plasma density and temperature; thus, it is different for tenuous and dense flares and may be different for flares with low and high spectral peak frequencies. 
We expect that the study of X-ray emission which we plan to perform in a follow-up paper will allow us to better constrain the causes of high acceleration efficiency and low  direct heating in the \csf s. 

The cause of the hard-soft-hard (HSH) evolution pattern of the high-frequency spectral index $\alpha_{\rm hf}$, which was found to occur in $\sim$40\,\% of cold flares (see Section~\ref{ssec_analysis_mw_ev}), is unclear. 
To draw reliable conclusions about the reasons responsible for the evolution of $\alpha_{\rm hf}$ one would have to perform thorough case studies of well-observed flares. 
These studies seem to be promising as evolution patterns of $\alpha_{\rm hf}$ might not be features solely distinctive of \csf s, but may be common to other solar flares accompanied by electron acceleration. 

In the forthcoming part of this study we will investigate energy partitioning between thermal and nonthermal flare components based on X-ray observations to verify if there was truly no (or very little) direct heating for the flares from the present list.
Analysis of hard X-ray data will allow recovery of the spectrum of accelerated electrons more directly and, thus,  help disentangling acceleration, propagation, and other effects on the  electron spectra and the high-frequency spectral index of the radio emission. 

\newpage
\section{Conclusions}

This study identifies and statistically analyzes about 100  ``cold'' solar flares, with weak thermal response in the soft X-ray range relative to the prominent nonthermal emission in the \mw\ range. 
The statistical analysis of these cold flares in the \mw\ range confirms the conclusions obtained in the previous study by \cite{Lysenko2018}:
\csf s are characterized by higher peak frequencies of \gyr\ emission, harder spectral indices in the high frequency range, and shorter durations than the flares of the reference group taken from the statistical study by \cite{Nita2004}. 
This study reveals that, for a majority of cold flares, the \gyr\ spectrum is influenced by the Razin effect rather than \gyr\ self-absorption. 
We propose that the majority of cold flares are confined flares and are associated with short loops with strong magnetic fields and dense ambient plasma. 
However, cold flares do not represent a homogeneous group, and there are cold flares  with low or moderate values of peak frequencies and long duration. 
We suggest that in \csf s the direct plasma heating is negligible and almost all the heating is driven by the Coulomb energy loss of the accelerated electrons. 
A better understanding of why the thermal emission is weak in \csf s requires an analysis of the X-ray emission, which will be performed in a subsequent study.

\acknowledgements
We thank Dr. Dmitry Svinkin for useful advice concerning flare selection and data analysis.
The work of A.L.L. was carried out in the framework of the basic funding program of the Ioffe Institute No. 0040-2019-0025.
G.D.F. was supported 
by NSF grants AGS-2121632,  
and AST-2206424,  
and NASA grants
80NSSC19K0068, 
and 80NSSC23K0090, 
to New Jersey Institute of Technology. 
D.A.Z., A.T.A. and N.S.M. acknowledges the support of the Ministry of Science and Higher Education of the Russian Federation. 
The BBMS data were obtained using the Unique Research Facility Siberian Solar Radio Telescope\footnote{\url{http://ckp-angara.iszf.irk.ru/index_en.html}}. 
We are grateful to the teams of the RSTN, Nobeyama Radio Observatory, SDO and RHESSI, who have provided open access to their data. 
G.G.M. was supported by grant 21-16508J of the Grant Agency of the Czech Republic, the project RVO:67985815, and the project LM2018106 of the Ministry of Education, Youth and Sports of the Czech Republic. 
S.W. acknowledges support from AFOSR under grant 23RVCOR003.

\facilities{Nobeyama Radio Observatory (NoRP and NoRH), RSTN, SDO/AIA}

\bibliography{ColdFlareStatMW.bib}

\end{document}